\documentclass[aps,prl,twocolumn,superscriptaddress,nofootinbib]{revtex4-1}

\usepackage[colorlinks=true,citecolor=red,urlcolor=blue]{hyperref}
\usepackage{amssymb}
\usepackage{amsmath,color}
\pdfoutput=1
\usepackage{graphicx}
\usepackage{bbm}
\usepackage{verbatim}
\usepackage[T1]{fontenc}
\usepackage[utf8]{inputenc}
\usepackage[normalem]{ulem}

\usepackage{url}
\usepackage{xcolor}
\usepackage{caption}
\usepackage{subcaption}

\usepackage{diagbox}
\usepackage{color}

\usepackage{multirow}
\usepackage{hhline}
\usepackage{tabularx}

\begin{document}


\title{The detection, extraction and parameter estimation of extreme-mass-ratio inspirals with deep learning}


\author{Qianyun Yun}
\email[]{yqy@shao.ac.cn}

\affiliation{Shanghai Astronomical Observatory,  Chinese Academy of Sciences,  Shanghai,  China,  200030}
\affiliation{Hangzhou Institute for Advanced Study, University of Chinese Academy of Sciences, Hangzhou 310124, China}

\affiliation{School of Physics and Astronomy, Shanghai Jiao Tong University 800 Dongchuan RD.,Minhang District, Shanghai, 200240, China}

\author{Wen-Biao Han}
\email{Corresponding author: wbhan@shao.ac.cn}
\affiliation{Hangzhou Institute for Advanced Study, University of Chinese Academy of Sciences, Hangzhou 310124, China}
\affiliation{Shanghai Astronomical Observatory,  Chinese Academy of Sciences,  Shanghai,  China,  200030}

\affiliation{School of Astronomy and Space Science,  University of Chinese Academy of Sciences,  Beijing,  China,  100049}
\affiliation{Taiji Laboratory for Gravitational Wave Universe (Beijing/Hangzhou), University of Chinese Academy of Sciences, Beijing 100049, China}

\author{Yi-Yang Guo}
\affiliation{Lanzhou Center for Theoretical Physics, Key Laboratory of Theoretical Physics of Gansu Province,\\
and Key Laboratory of Quantum Theory and Applications of MoE,\\
Lanzhou University, Lanzhou, Gansu 730000, China
}
\affiliation{Institute of Theoretical Physics \& Research Center of Gravitation, Lanzhou University, Lanzhou 730000, China}

\author{He Wang}
\affiliation{International Centre for Theoretical Physics Asia-Pacific (ICTP-AP), University of Chinese Academy of Sciences (UCAS), Beijing, China
}
\affiliation{Taiji Laboratory for Gravitational Wave Universe (Beijing/Hangzhou), University of Chinese Academy of Sciences, Beijing 100049, China}

\author{Minghui Du}
\affiliation{ Center for Gravitational Wave Experiment, National Microgravity Laboratory, Institute of Mechanics, Chinese Academy of Sciences, Beĳing 100190, China}

\date{\today}

\begin{abstract}
One of the primary goals of space-borne gravitational wave detectors is to detect and analyze extreme-mass-ratio inspirals (EMRIs). This endeavor presents a significant challenge due to the complex and lengthy EMRI signals, further compounded by their inherently faint nature. In this letter, we introduce a 2-layer Convolutional Neural Network (CNN) approach to detect EMRI signals for space-borne detectors, achieving a true positive rate (TPR) of 96.9~\% at a 1~\% false positive rate (FPR) for signal-to-noise ratio (SNR) from 50 to 100.  Especially, the key intrinsic parameters of EMRIs such as the mass and spin  of the supermassive black hole~(SMBH) and the initial eccentricity of the orbit can be inferred directly by employing a VGG network. The mass and spin of the SMBH can be determined at 99~\% and 92~\% respectively. This will greatly reduce the parameter spaces and computing cost for the following Bayesian parameter estimation. Our model also has a low dependency on the accuracy of the waveform model. This study underscores the potential of deep learning methods in EMRI data analysis, enabling the rapid detection of EMRI signals and efficient parameter estimation.
\end{abstract}

\maketitle

\textit{Introduction.} Since the first detection of GW150914~\cite{LIGOScientific:2016dsl}, ground-based gravitational wave (GW) detectors such as LIGO~\cite{Harry:2010zz} and VIRGO~\cite{VIRGO:2014yos}  have made remarkable progress. These detectors have subsequently detected nearly one hundred GW events of the merger of two black holes with stellar mass before 2022~\cite{LIGOScientific:2021djp}. These observations  enable researchers to test Einstein's general theory of relativity~\cite{LIGOScientific:2019fpa} and explore the universe through an independent approach~\cite{LIGOScientific:2021aug}.

Ground-based detectors such as LIGO, Virgo, and Kagra~\cite{KAGRA:2013rdx} focus on gravitational waves within a frequency range of 10 Hz to a few thousand Hertz, as determined by their sensitivity curves~\cite{LIGOScientific:2021djp, Kaiser:2020tlg}. While space-borne  interferometers, represented by missions like LISA~\cite{LISA:2017pwj} and Taiji~\cite{Hu:2017mde} are specifically engineered to detect low-frequency gravitational waves, covering the frequency from 0.1 mHz to 1 Hz. One primary objective of these space-borne interferometers is to detect extreme-mass-ratio inspiral systems~\cite{Babak:2017tow, Ruan:2018tsw, Han:2018hby}. 


An EMRI system consists of a stellar-mass black hole orbiting around a much heavier black hole with a mass ranging from $10^4-10^7M_{\odot}$.  These supermassive black holes (SMBHs) are typically located at the centers of  galaxies~\cite{Babak:2017tow}. The inspirals of compact objects around the SMBHs emit  GWs at low frequencies. Investigating EMRI systems provides opportunities to examine theories of gravity~\cite{Babak:2017tow, Han:2018hby}. This exploration enables precise mapping  mapping of the spacetime around  SMBHs and the testing of the no-hair theory~\cite{Maselli:2021men, LISA:2022kgy}.  Moreover, it can help us  comprehend the mass distribution of SMBHs and their relationships with host galaxies, as well as  trace the formation and evolution of them~\cite{LISA:2022yao}.

Analyzing EMRI signals  is very challenging due to their prolonged duration and intricate characteristics. These signals can span  for several years, demanding massive computational resources to generate sufficient and high accurate waveforms for data analysis~\cite{Cornish:2008zd, Babak:2009ua}. Traditional gravitational wave data analysis methods, like match filtering and Bayesian parameter estimation, require a minimum of $10^{40}$ EMRI waveform templates to cover the 14-parameter space~\cite{Gair:2004iv}. Even with the application of the F–statistic algorithm~\cite{Wang:2012xh}, a substantial amount of templates are still required. In addition, the traditional methods heavily depend on template accuracy in data analysis. Nevertheless, the intricate harmonics and modulations of EMRI signals can create difficulties in modeling an accurate template due to factors like self-force effects.

Deep learning  techniques have demonstrated their remarkable efficacy in GW data analysis. They excel in both detecting and characterizing GW signals for both simulated and real situations~\cite{George:2017pmj, Wang:2019zaj, Jin:2023ahl}. Additionally, deep learning has already been used for the detection of EMRIs~\cite{Zhang:2022xuq, Zhao:2022qob}. Moreover, deep learning serves as a valuable tool for estimating galaxy parameters~\cite{tuccillo2018deep}, which can also be employed for GW parameter estimation.

In this letter, we use a 2-layer CNN network designed for detecting EMRI signals in the time-frequency domain. We compute the EMRI signals using several waveform models and simulate the noise using Taiji's analytical PSD. Considering the practical application in space detectors, we incorporate second-generation Time Delay Interferometry (TDI)~\cite{Amaro-Seoane:2012vvq}. Then we apply the Q-transform to convert the time series data into time-frequency data before feeding them into the network.  We also employ a UNet network to extract EMRI signals from noise and estimate their parameters using a VGG net. Though our research primarily concentrates on Taiji data, our method is versatile for other space-borne GW detectors like LISA.

\begin{figure}
\includegraphics[width=0.5\textwidth]{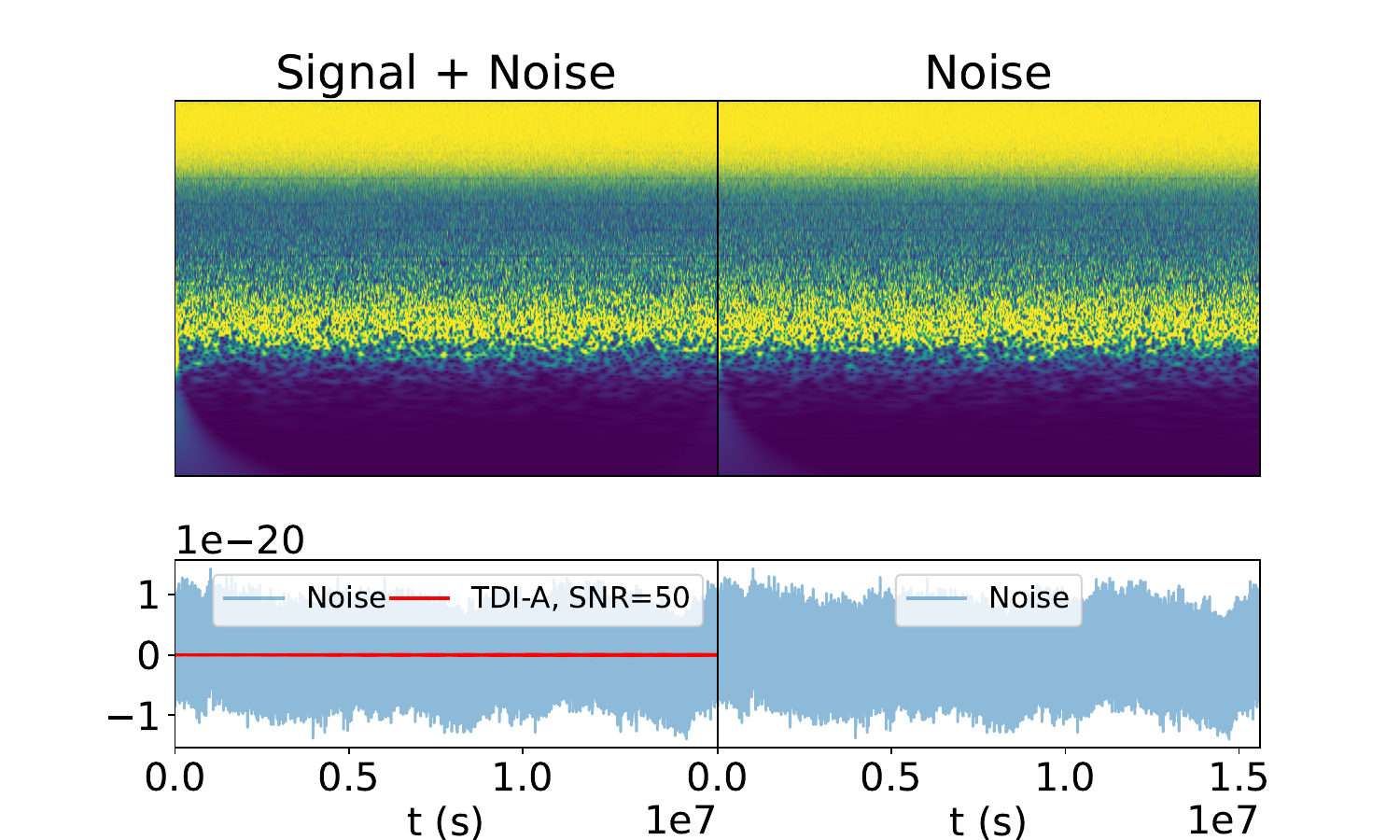} 

\caption{\textbf{Left:} The CQT plot and the time series plot of signal and noise. \textbf{Right:} The CQT plot and the time series plot of pure noise. }
\label{fig:Training_data}
\end{figure}

\textit{Methodology.} The initial step in training a deep learning Network is the creation of the training and testing dataset. Our dataset consists of two elements: the combination of signals and noise ($d = h + n$), labeled as 1, and noise alone ($n$), labeled as 0. The demonstration of these two types of data is shown in Fig.~\ref{fig:Training_data}.

We employ the AAK model~\cite{Chua:2017ujo} within specified parameter ranges (see Table~\ref{paramters}) to generate EMRI signals. These signals last 0.5 years and are sampled at 10-second intervals. We utilize second-generation TDI techniques~\cite{Amaro-Seoane:2012vvq} to get the A, E, and T channel data of the Taiji detector. TDI is a crucial method for reducing laser frequency noise and ensuring desired sensitivity in space-borne detector missions. We chose the A and E channels in this letter. Subsequently, we compute SNR values using the analytical PSD, maintaining them within the range of 50-100. The mathematical expression for the PSD of the X channel ($ A = \frac{1}{\sqrt{2}}(Z-X), E = \frac{1}{\sqrt{6}}(X-2Y+Z)$)  is as follows~\cite{Ren:2023yec}: $\mathrm{PSD}{X_2} = 64 \sin^2(\omega L) \sin^2(2 \omega L) \left(P_{\mathrm{oms}} + (3 + \cos(2 \omega L)) P_{\mathrm{acc}}\right)$, where $\omega = 2 \pi f / c$, and $P_{\mathrm{oms}}(f)$ represents high-frequency noise from the optical metrology system, while $P_{\mathrm{acc}}(f)$ results from test mass acceleration. We utilize this PSD to create the detector's noise. Half of  these noises are combined with the signals, as illustrated in the bottom-left panel of Fig.\ref{fig:Training_data}.  The data of pure noise is displayed in the bottom-right panel of Fig.\ref{fig:Training_data}.  The dataset consists of 6000 samples for both $d$ and $n$.  The dataset is subsequently divided into training and testing sets, with the former for model training and the latter for evaluation.
\begin{table}[ht]
\centering
\renewcommand{\arraystretch}{1.6}
\caption{The parameter ranges of AKK and effective-one-body-Teukolsky waveforms for training and testing data.}
\small
\renewcommand {\arraystretch} {1}
\begin{tabular}{p{1.5cm}|p{2.5cm}|p{1.5cm}|p{2.5cm}}
\hline\hline
Parameter & Range (Uniform distribution)&Parameter & Range (Uniform distribution)\\
\hline
$M/M_{\odot}$ & ($10^5, 10^8$)&$\gamma$ & (0, 0.1) \\
$\mu/M_{\odot}$ & (10, 100) ; $M/\mu \geq 10^4 $&$\theta_S$ & (0, $\pi$)\\
$a$ & $(10^{-3}$, 0.90)&$\phi_S$ & (0, $2\pi$ ) \\
$e_0$ & (0.005, 0.6)&$\theta_K$ & (0, $\pi$) \\
$p_0/M$ & (10, 12)&$\phi_K$ & (0, $2\pi$) \\
$\iota$ & (0, 0.1) & &  \\
 
\hline\hline
\end{tabular}
\label{paramters}
\end{table}

Time-frequency analysis proves to be a highly effective approach in EMRI data analysis~\cite{Wang:2012xh}. Dominant harmonics in EMRI signals become clear in the time-frequency domain. Therefore, we chose to employ the Constant Q Transform (CQT) for data preprocessing before feeding them into the networks.
CQT is a special case of Variable Q transform(VQT). Both of them are related to complex Morlet wavelet transform. The definition is:
   $\delta f_k=2^{1/n}\cdot\delta f_{k-1}=\left(2^{1/n}\right)^k\cdot\delta f_{\min}$,
where $\delta f_k$ is the bandwidth of the k-th filter, $f_{min}$ is the central frequency of the lowest filter, and n is the number of filters per octave. Define the quality factor Q for  CQT:
$ Q=\frac{f_k}{\delta f_k}$.
With this factor, we can define window length: $ N[k]=\frac{f_{s}}{\delta f_{k}}=\frac{f_{s}}{f_{k}}Q$. Hamming window is one of the most commonly used windows, take this for example: $ W[k,n]=\alpha-(1-\alpha)\cos\frac{2\pi n}{N[k]-1},\quad\alpha=25/46,\quad0\leqslant n\leqslant N[k]-1$. With above equations, define Q transform as: $X[k]=\frac{1}{N[k]}\sum_{n=0}^{N[k]-1}W[k,n]x[n]e^{\frac{-j2\pi Qn}{N[k]}}$.
We use \textbf{librosa}~(https://librosa.org/doc/latest/index.html) to do the process, store the Q-transform results as gray images, and resize them to (255,255) matrixes as the input data. The top panel of Fig.~\ref{fig:Training_data} shows the training data post-Q-transform, where the input signal has an SNR of 50.  In this figure, EMRI features are not distinctly visible due to noise, making it challenging to visually distinguish between $d$ and $n$. To solve this problem, we try to use CNN  to distinguish these two kind of data. 

For this target, we adopt a basic framework of a CNN network architecture, which includes two Conv2d layers with filter sizes of 16 and 32.  The convolutional layers have kernel sizes of $3 \times 3$ and $3 \times 3$, respectively. Following each convolutional layer, a Batchnorm layer is introduced, followed by the application of a standard ReLU activation function as non-linearities layers, and subsequently, a max-pooling layer is applied.  The convolution layers have a stride of 1, while the pooling layers have a stride of 2. Following the two convolutional blocks, there are  two fully connected layers with sizes of 128 and 64.  

Then we use the simulated data to train our detection network. During the training, the cross-entropy loss is employed as the loss function, defined as:
\begin{equation}
    L = \sum_{i=1}^{N} \left( -y_{i_{\text{true}}} \cdot \log(y_{i_{\text{pred}}}) - (1 - y_{i_{\text{true}}}) \cdot \log(1 - y_{i_{\text{pred}}}) \right).
\end{equation}  The loss quantifies the difference between our model's predictions and the true labels. Through backpropagation, the model updates each parameter to minimize the loss. The training goes through  200 epochs. The Adam optimizer is employed with a learning rate set at 0.01.

 The performance of the network is assessed by testing data and the results  are visually shown by the Receiver operating characteristics (ROC) curve. The ROC shows our model's ability to  distinguish $d$ and $n$ by plotting the true positive rate against the false positive rate.  The quantified performance is determined by the Area Under the ROC Curve (AUC). A higher AUC score, closer to 1, means that the model has a better capability to  distinguish the EMRI signals. 

\begin{figure}
\includegraphics[width=0.45\textwidth]{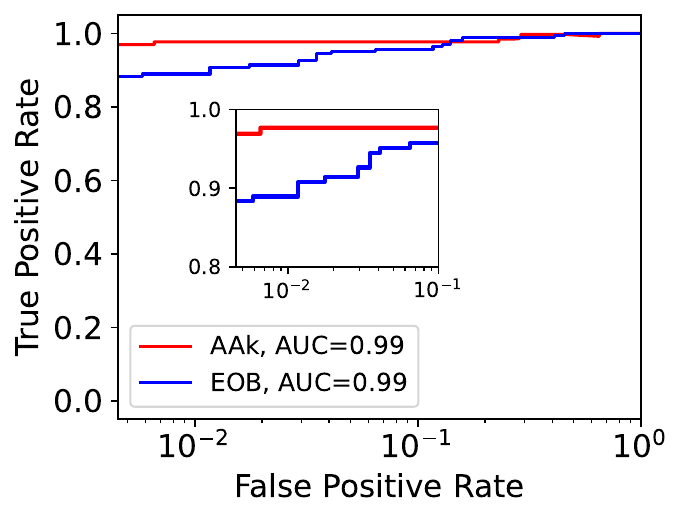} 

\caption{ROC curves for the trained CNN model  to detect EMRI signals generated by the AAK (red) and EOB (blue) waveform model. The SNRs of the testing data range in [50,100]. The figure inside zooms in on the TPR from 0 to 0.1. }
\label{fig:roc}
\end{figure}

 In Fig.~\ref{fig:roc}, the red curve shows the performance of  our model in detecting EMRI signals simulated by the AAK templates, achieving an AUC of 99~\%. The detection achieves a 96.9\% TPR while  1\% FPR with SNRs from 50 to 100. 
 
 To evaluate the robustness of our network, we also utilize an alternative waveform model to generate test data. This is crucial since the theoretical  EMRI signals may differ from the true EMRI signals due to the accuracy and the real scenarios. Traditional GW detection relies heavily on accurate waveform templates,  which may lead to missed detections when true signals differ from the theoretical ones. We believe that deep learning can reduce this template dependency. To confirm this, the effective-one-body-Teukolsky (ET)~\cite{Shen:2023pje} waveform model is used to generate new data to test our detection network trained with AAK waveforms.

In Fig.~\ref{fig:roc}, the blue ROC curve shows the capability to detect EMRI signals generated by ET model. The CNN network also performs effectively on this new test dataset, with an AUC of 99~\%. The TPR can be still 90\% at 1\% FPR. This result shows the  robustness of our network in the EMRI detention.  It indicates that even if the theoretical waveforms are not accurate enough, our network retains the capability to identify the EMRI signals.

\begin{figure}
\includegraphics[width=0.5\textwidth,height=0.12\textheight]{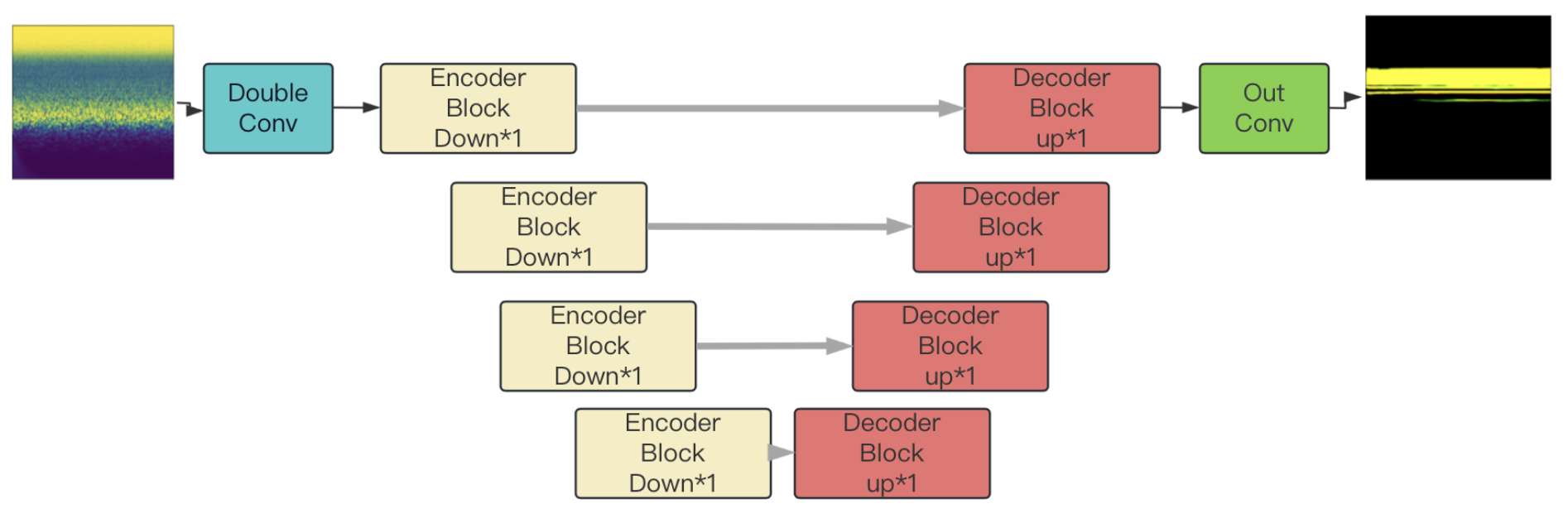} 
\includegraphics[width=0.45\textwidth]{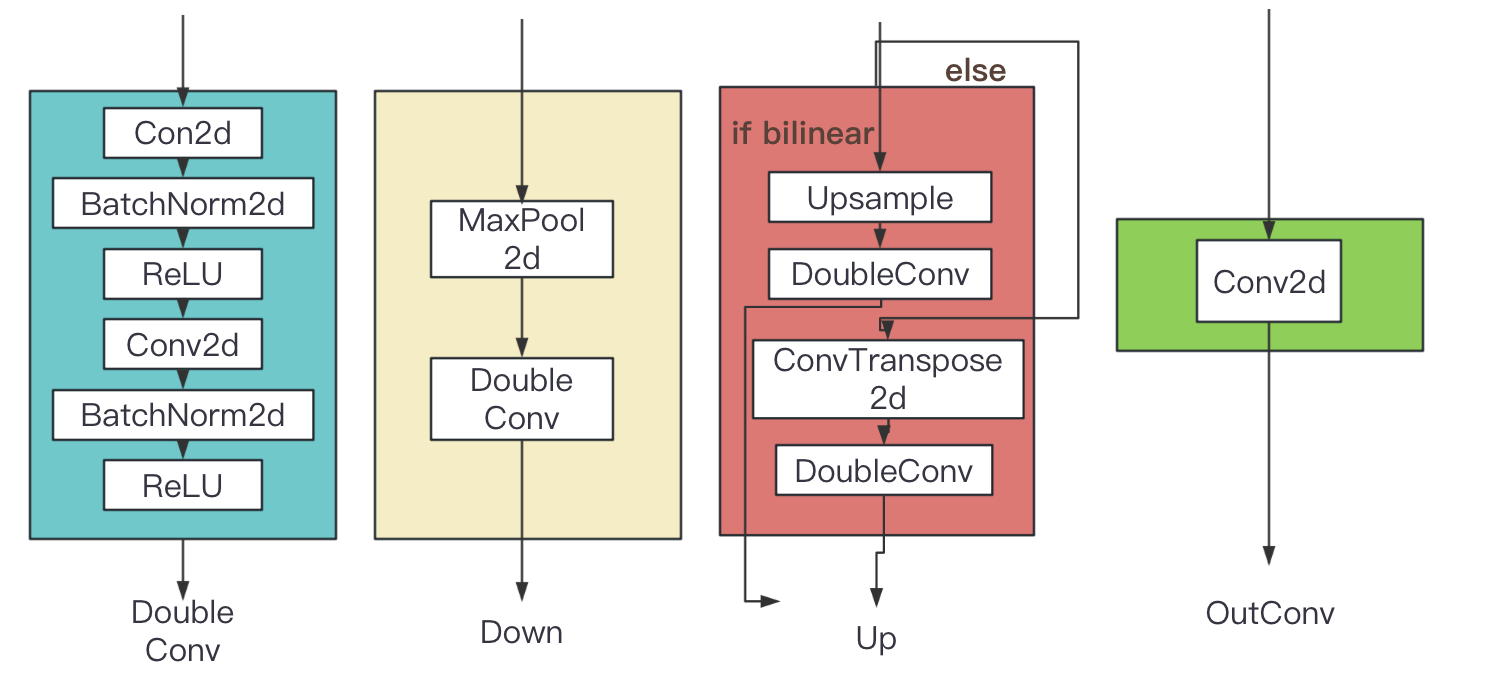}  

\caption{\textbf{Top:} Structural details of our UNet network in this letter. \textbf{Bottom:} The details of each module for our UNet network in this letter. }
\label{fig:Unet}
\end{figure}

After successfully recognizing EMRI signals,  we employ the UNet network to extract them~\cite{ronneberger2015u}. The extraction process utilizes signal-with-noise as input data and pure signals as the target~(https://github.com/milesial/Pytorch-UNet/tree/master/unet).  This Unet network's architecture is illustrated in Fig.~\ref{fig:Unet}, which depicts the process of extracting the final time-frequency signal from the original image.

UNet comprises an encoder and a decoder. The encoder is composed of four ``Down'' blocks, reducing the input image's resolution and extracting the crucial features. The decoder, composed of four ``Up'' blocks, performs upsampling to restore features to the input size, and then generates the final output. ``Skip Connections'' is the key process, which can merge features of the encoder and decoder at the same hierarchical level~(four gray parallel lines with arrows in the top panel of Fig.~\ref{fig:Unet}). Details of the blocks are illustrated at the bottom panel of Fig.~\ref{fig:Unet}.

 During the training, we utilize Binary Cross-Entropy with Logits as the loss function: 
 \begin{equation}
     \text{L} = \text{max}(y_{\text{pred}}, 0) 
- y_{\text{pred}} \cdot y_{\text{true}} + \log(1 + \exp(-\text{abs}(y_{\text{pred}}))).
 \end{equation} This loss function is commonly utilized in Unet-based segmentation tasks to evaluate the difference between predicted outputs and actual targets. We train 50 epochs to improve the quality of the output, and Fig.~\ref{fig:Unet} shows how we extract a time-frequency EMRI signal from a noisy one using the Unet network.

\begin{figure}
\includegraphics[width=0.5\textwidth]{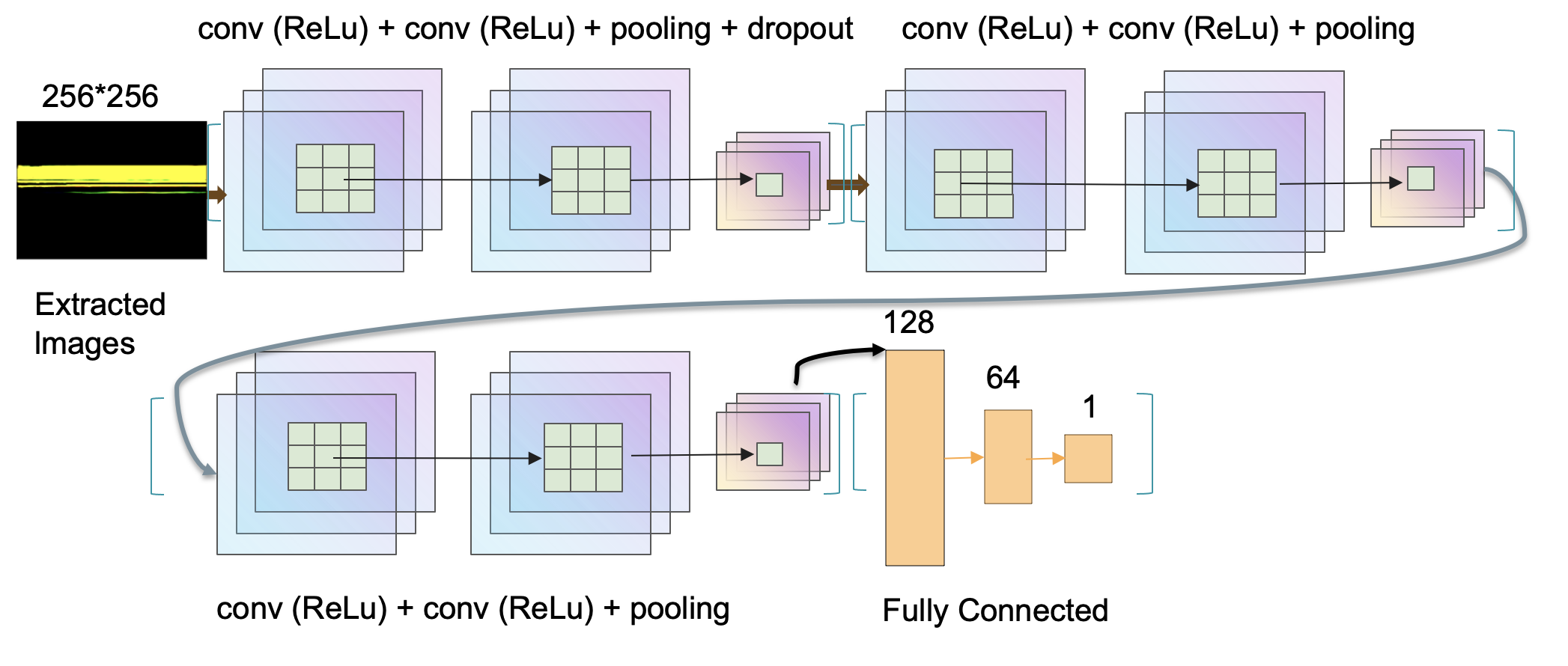} 

\caption{The scheme of Architecture in the parameter estimation work.}
\label{fig:VGG}
\end{figure}

After the extraction, we use a VGG~\cite{simonyan2014very} network for parameter estimation, inspired by the approach in Ref.~\cite{tuccillo2018deep}, where deep learning was applied for estimating galaxy parameters. The architecture of our model is depicted in Fig.~\ref{fig:VGG}. In this model, input images go through three blocks. The initial block consists of two 2D convolution layers with a 4x4 filter, followed by a 2x2 max pooling layer and a dropout layer. The subsequent two blocks have the same architecture, each comprising 4x4 convolution and max pooling layers.

\begin{figure}
\includegraphics[width=0.35\textwidth]{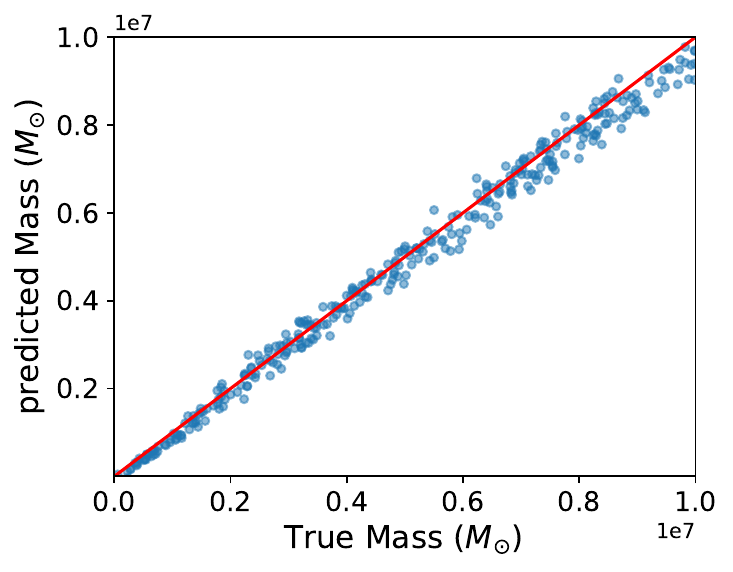} 
\includegraphics[width=0.345\textwidth]{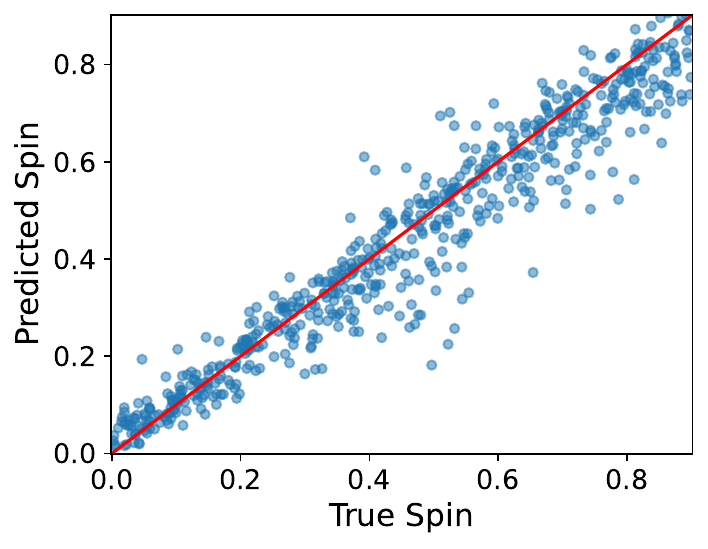} 
\includegraphics[width=0.355\textwidth]{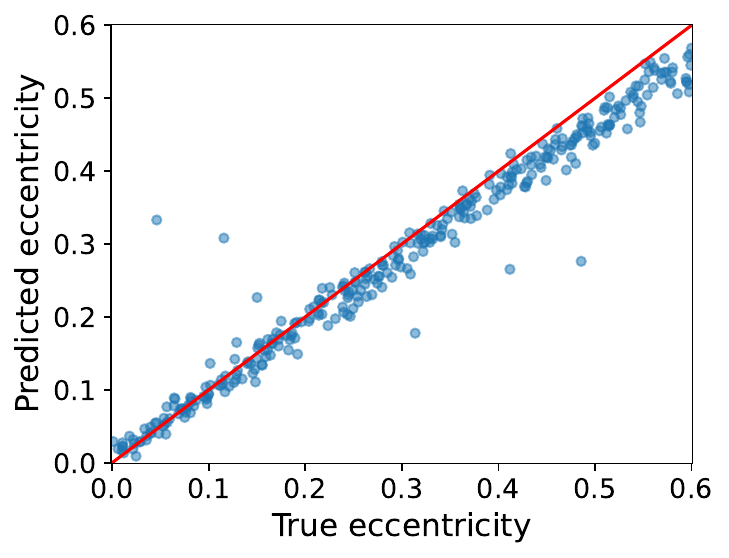} 
\caption{The parameter estimation results from the VGG net of the mass values, spin values, and initial eccentricity values of the center SMBHs. }
\label{fig:parameter result}
\end{figure}

We then begin to train the VGG network for parameter estimation. Firstly, we create a new dataset with signals and noise, dividing it into training and testing sets. Following this, we again use the previous UNet model to extract signals from the noise and input these extracted signals to the VGG network. The true parameters of the signals are the targets. The network undergoes 50 epochs of training. We utilize Mean Squared Error (MSE) as the loss function, defined as: 
\begin{equation}
   \mathrm{L}=\frac{1}{N} \sum_{i=1}^N(y_{i_{\text{pred}}}-y_{i_{\text{true}}})^2 .
\end{equation}After training, we input testing data to extract signals and obtain their parameters. Fig.~\ref{fig:parameter result} illustrates the relationship between our predicted values and the true values.

In the top panel of Fig.~\ref{fig:parameter result}, each point represents the true and predicted mass of the SMBH in an EMRI. The accuracy of our predictions is calculated by the coefficient of determination $R^2$, defined as $R^2=1-\frac{\sum_i^n\left(y_i-f_i\right)^2}{\sum_i^n\left(y_i-\bar{y}\right)^2}$. Here, $f_i$ denotes the predicted value, $y_i$ is the true value, and $\bar{y}$ is the mean of the true value. 
 Remarkably,  the predicted mass achieves an exceptionally good $R^2$ value of 0.989.  In the middle panel of Fig.~\ref{fig:parameter result},  we present predictions for the spins of SMBHs, yielding an $R^2$ of 0.92. Lastly, the bottom panel of Fig.~\ref{fig:parameter result} showcases our predictions for the initial eccentricity of the orbits, accompanied by an $R^2$ of 0.95.

\textit{Conclusion and discussion.}
Detection and parameter estimation of EMRIs are still a challenge for space-borne projects. Due to the weakness, complexity (at least 14 parameters), and long duration of the EMRI signals, the traditional matched-filtering and Bayesian inference may have difficulty in detecting EMRIs and obtaining accurate source parameters. The PE based on the Bayesian method requires very accurate waveform templates and huge amounts of computing resources. Fortunately, the booming machine learning technologies offer a new chance to overcome this problem.   
 In this Letter, our CNN-based model exhibits excellent performance for the detection of EMRIs and achieves a 96.9\% TPR at a 1\% FPR with SNRs from 50 to 100. Especially, our model also shows good robustness, if the training and testing data come from different waveform models, the TPR can be still 91\% at 1\% FPR. 

Furthermore, we also employ UNet to extract EMRI signals from noise and VGG to estimate their main intrinsic parameters, such as the mass and  spin of the SMBH. The statistical error is approximately 1~\% for mass, less than 10~\% for spin, and less than 5~\% for orbital eccentricity. These intrinsic parameters directly determine the physical characteristics of the EMRI sources, and will greatly reduce the parameter space and the computing cost for further Bayesian PE with Markov Chain Monte Carlo. 

In this letter, we demonstrate the great potential of deep learning in the area of detection, extraction, and parameter estimation of EMRIs.  However, in the present work, the SNRs of the signals are all larger than 50. Therefore, we will analyze the weaker EMRI signals in the upcoming work, with SNR as low as 20. This kind of signals are more popular for space-borne interferometers and will improve the detection rate.   

\begin{acknowledgments}
We would like to express our sincere appreciation to Quanfeng Xu~(SHAO) and Yuewei Zhang~(SJTU) for their invaluable assistance and guidance about the deep learning part throughout this research project. Thanks to the Lanzhou University Summer School on Gravitational Waves where we talk about the advantages of using time-frequency domain data to analysis EMRI signals with Xiaobo Zou (LZU) and discuss the CQT with Professor Soumya D. Mohanty~(The University of Texas Rio Grande Valley), lecturer Qunyin Xie (LZU), and her student Shuzhu Jin. This work was supported by the National Key R\&D Program of China (Grant Nos. 2021YFC2203002), the National Natural Science Foundation of China (Grant Nos. 12173071). Wen-Biao Han was supported by the CAS Project for Young Scientists in Basic Research (Grant No. YSBR-006). This work made use of the High Performance Computing Resource in the Core Facility for Advanced Research Computing at Shanghai Astronomical Observatory.
\end{acknowledgments}


\end{document}